\documentclass[11pt,a4paper]{article}
\usepackage[hyperref]{emnlp-ijcnlp-2019}
\usepackage{times}
\usepackage{latexsym}
\usepackage{url}
\usepackage{graphicx}
\usepackage{amsmath}
\usepackage{bm}
\usepackage[group-separator={,},group-minimum-digits={3}]{siunitx}
\usepackage{subfigure}
\usepackage{amssymb}
\usepackage{multirow}
\usepackage{xcolor}
\usepackage{enumitem}

\setitemize[1]{itemsep=3pt,partopsep=3pt,parsep=\parskip,topsep=3pt}

\title{Incorporating Fine-grained Events in Stock Movement Prediction }

\date{}
\author{Deli Chen\textsuperscript{1}\thanks{~This work is done when Deli Chen is a intern at Mizuho Securities.},
Yanyan Zou\textsuperscript{2},
Keiko Harimoto\textsuperscript{3},
Ruihan Bao\textsuperscript{3},
\textbf{Xuancheng Ren\textsuperscript{1},
Xu Sun\textsuperscript{1}}
\\ \\
\textsuperscript{1}{MOE Key Lab of Computational Linguistics, School of EECS, Peking University}\\
\textsuperscript{2}{StatNLP Research Group, Singapore University of Technology and Design}\\
\textsuperscript{3}{Mizuho Securities Co., Ltd.}\\
\{chendeli,renxc,xusun\}@pku.edu.cn,
yanyan\_zou@mymail.sutd.edu.sg,\\
\{keiko.harimoto,ruihan.bao\}@mizuho-sc.com 
}

\hypersetup{draft}
\begin{document}
\maketitle
\begin{abstract}
Considering event structure information has proven helpful in text-based stock movement prediction. However, existing works mainly adopt the coarse-grained events, which loses the specific semantic information of diverse event types. In this work, we propose to incorporate the fine-grained events in stock movement prediction. Firstly, we propose a professional finance event dictionary built by domain experts and use it to extract fine-grained events automatically from finance news. Then we design a neural model to combine finance news with fine-grained event structure and stock trade data to predict the stock movement. Besides, in order to improve the generalizability of the proposed method, we design an advanced model that uses the extracted fine-grained events as the distant supervised label to train a multi-task framework of event extraction and stock prediction. The experimental results show that our method outperforms all the baselines and has good generalizability.  
\end{abstract}

\section{Introduction}
Stock movement plays an important role in economic activities, so the prediction of stock movement has caught a lot of attention of researchers. In recent years, employing the stock related text (such as finance news or tweets) has become the mainstream~\citep{Raw2Emnlp14sentiment,DingEvent2,Raw3MultiModal2AAAI15TensorMethod,StockOther1,StockOther3,StockOther4} of stock movement prediction task.
In these text-based stock prediction works, various methods are proposed to extract semantic information from stock related text to help the prediction of stock movement.
There are mainly two methods of applying text: employing raw text~\citep{HANN,Raw4ACL18VAE} or coarse-grained $<$S,P,O$>$ structure (subject, predicate and object) extracted from text~\citep{DingEvent3,Event4MultiModal1IEEE18MultiSource}. In the previous studies, the latter method has proven more powerful than the former one, which demonstrates that the event structure containing semantic information is helpful for stock movement prediction. 
\begin{figure}[t]
\centering
\includegraphics[scale=0.35]{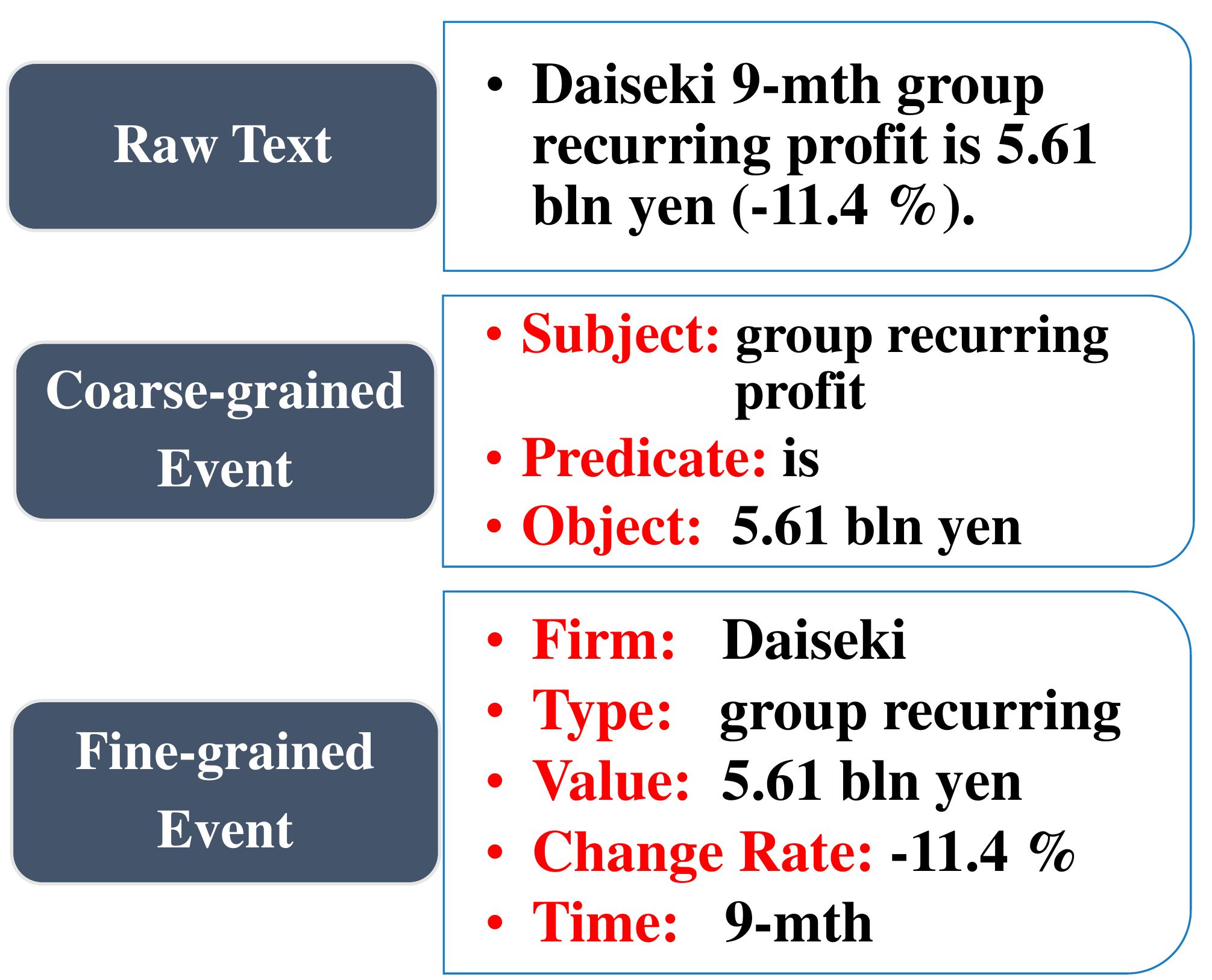}
\caption{The same news of \textit{Earnings Profit} event in different forms. The event structure consists of event roles (red words) which are the key point of the semantic information.}
\label{fig0sample}
\vspace{-0.2in}
\end{figure}
Figure~\ref{fig0sample} shows a piece of news of \textit{Earnings Profit} event in different forms: raw text, coarse-grained event ($<$S,P,O$>$) and fine-grained event~\citep{ChineseFinancialNewsEventExtraction,AttGraphMultiEE}. We observe that there are still some issues with the $<$S,P,O$>$ method. Firstly, the $<$S,P,O$>$ method only extracts subject, predicate and object, which misses some important event roles, such as the earnings \textit{Time} and \textit{Change Rate}, which are included in the fine-grained event. Besides, applying $<$S,P,O$>$ structure for all event types loses the specific semantic structure in different types of finance events. In Figure~\ref{fig0sample}, the fine-grained event employs \textit{Type} instead of \textit{Subject} used in the coarse-grained event and employs \textit{Value} instead of \textit{Object}, which can describe the event roles in a more detailed way. 
In this work, we propose to incorporate the fine-grained events in one-day-ahead stock movement prediction. The fine-grained event structure describes the specific framework and key points of various finance events. Applying fine-grained events is beneficial for learning a better text representation because the finance knowledge contained in event structure is helpful for understanding the semantic information. 

Inspired by the automatic event data generation method~\citep{LargeEventExtraction,ScaleUPEE,ChineseFinancialNewsEventExtraction}, we propose the TOPIX\footnote{Tokyo Stock Price Index, commonly known as TOPIX, is an important stock market index for the Tokyo Stock Exchange (TSE) in Japan.} Finance Event Dictionary (\textbf{TFED}) built by domain experts with professional finance knowledge and adopt it to extract fine-grained events automatically for most of finance news. Then we design two different neural models: Structured Stock Prediction Model (\textbf{SSPM}) and Multi-task Structured Stock Prediction Model (\textbf{MSSPM}). SSPM fuses the extracted fine-grained event and news text firstly, and then conduct interaction between text data and stock trade data to make prediction. SSPM outperforms all the baselines but it can hardly handle the news that can not be recognized by TFED, which we call uncovered news, so MSSPM is designed to learn event extraction using the fine-grained events as the distant supervised label. Besides, we propose to learn event extraction and stock prediction jointly in MSSPM because these two tasks are highly related. The improvement of event extraction result can boost news understanding and promote the stock prediction. And the output of stock prediction can give feedback to event extraction. So the joint learning can share valuable information between tasks. Result shows that MSSPM outperforms SSPM on the uncovered news and increases the method's generalizability.
The contributions of this work are summarized as follows:
\begin{itemize}
\item We propose to incorporate the fine-grained events in stock movement prediction and this method outperforms all the baselines.
\item We propose to learn event extraction and stock prediction jointly, which improves the method generalizability for uncovered news.
\item We propose TFED and a pipeline method which can extract fine-grained events from finance news automatically.
\item We propose the embedding method for minute-level stock trade data, and adopt time-series models to learn its representation.
\end{itemize}

\section{Related Work}
\subsection{Automatically Event Data Labeling}
 According to~\citep{DynamicMultiPoolingCNN,AttGraphMultiEE,ZeroShotEE}, the fine-grained event structure contains event types, event trigger words and event roles.  
\citet{TwitterEvent} propose a framework to extract events from twitter automatically.
~\citet{ChineseFinancialNewsEventExtraction} employ a predefined dictionary to label events and then extract document-level events from Chinese finance news. However, they only conduct experiments on 4 event types. While we employ a widely-covered dictionary with 32 different event types. ~\citet{LargeEventExtraction} adopt world and linguistic knowledge to detect event roles and trigger words from text.~\citet{ScaleUPEE} use the Freebase CVT structure to label data and extract event.~\citet{DSOpenEventExtraction} adopt distant supervision to extract event from open domain. There are some works using either manual rules~\citep{supply2_rule4event_detection} or machine learning methods~\citep{supply3_ml4event_detection} for finance event detection, while our event labeling method is stock specific with professional domain knowledge. 

\subsection{Stock Movement Prediction}
Many works using related text for stock movement prediction take the raw text as model input directly. \citet{Raw4ACL18VAE} adopt a variational generation model to combine tweets and stock history data to make the prediction.~\citet{Raw2Emnlp14sentiment} employ the sentiment analysis to help the decision.~\citet{Raw3MultiModal2AAAI15TensorMethod} adopt the tensor decompose method to get the interaction information of different inputs.~\citet{Raw1Coling18NewsBody} use the summary of news body instead of news headline to predict the stock returns.
Some other works try to employ structure information to predict the stock movement.~\citet{DingEvent1} extract $<$S,P,O$>$ (subject, predicate and object) structure from news to predict the stock movement. Then they propose two improved method based on $<$S,P,O$>$ structure by applying the weighted fusion of event roles ~\citep{DingEvent2} and introducing the entity relation knowledge~\citep{DingEvent3}. Besides,~\citet{Event4MultiModal1IEEE18MultiSource} employ a RBM to process $<$S,P,O$>$ to get the event representation.
\section{Fine-grained Event Extraction}
\subsection{TOPIX Finance Event Dictionary}
As shown in~\citep{ChineseFinancialNewsEventExtraction,ScaleUPEE} automatic fine-grained event extraction needs an event dictionary to define the event types. Each event type consists of event trigger words and event roles. News containing trigger words perhaps belongs to this event type. The event roles are the key points of semantic structure of this event type. However, there is no specific event dictionary for stock related finance events. So we hired three domain experts to summarize the high-frequency finance events which have a significant impact on stock trading and determine the event trigger words and event roles. 
With help of domain experts, we also annotated some auxiliary information for the following event extraction process: the POS label of the event roles, the dependency relation pattern of the event types and the necessary/unnecessary label of event roles. 
Not all event roles will appear in every instance of this event type. Take the \textit{Earnings Profit} event in Figure~\ref{fig0sample} for example, the \textit{Firm}, \textit{Type} and \textit{Value} will appear in every \textit{Earnings Profit} instance. But the \textit{Change Rate} and \textit{Time} may not appear in some \textit{Earnings Profit} instances. We regard the news containing all the necessary roles as an instance of related event.

TFED contains 32 types of finance events in 8 categories and covers all the main types of finance events that are highly related to stock movement, such as \textit{Earnings Profit}, \textit{M\&A} and \textit{Credit Ratings}. All the 32 event types of TFED are displayed in supplementary material A, as well as their trigger words and event roles.

\subsection{Event Extraction Process}
\label{Section2EventDataGenerate}
There are 4 steps in the event extraction process, in which we extract the fine-grained event structures from finance news. 
\paragraph{1. Extract Auxiliary Information.} In this step, we extract the auxiliary information of news: \textbf{POS Tagging} (lexical information) and \textbf{Dependency Relation} (syntactic information) by the popular Standford CoreNLP\footnote{\url{https://stanfordnlp.github.io/CoreNLP/}}~\citep{StandfordCoreNLP}. 

\paragraph{2. Filter Event Candidates.}We filter the news that may be an event instance by the TFED. News that contains any trigger word(s) in the dictionary will be regarded as a candidate of the related event. For example, the news in the Figure~\ref{fig0sample} is a candidate for the \textit{Earnings Profit} event because it contains the trigger word \textit{profit}. 

\paragraph{3. Locate Event Roles.}We regard news containing all the necessary event roles as an event instance. For event candidates driven by trigger words, we adopt matching rules set by domain experts to check the dependency relation and POS information. Firstly, we match the dependency relation of the candidate news with predefined dependency relation pattern of this event type in TFED to locate the event roles and check if all the necessary event roles are recalled. Then we check if all the event roles' POS labels are consistent with predefined labels. Only if these two conditions are satisfied, this news will be regarded as an event instance and the event roles are determined.

\paragraph{4. BIO Post-process.}The result of Step~3 is the label for event roles. Since we want to get the event label for each word in news, we use the BIO label standard to normalize the labeling result.
After all these 4 steps, we access the fine-grained event of news. And the extraction result shows that our method covers 71$\%$ news in the 210k samples, which proves that the TFED and the pipeline method work well on our experiment data. And for uncovered news, adding more event types is of high cost and low efficiency, so we extract the $<$S,P,O$>$ structure as replacement following the approach in ~\citep{Event4MultiModal1IEEE18MultiSource}.

\begin{figure*}[h]
\centering
\includegraphics[scale=0.43]{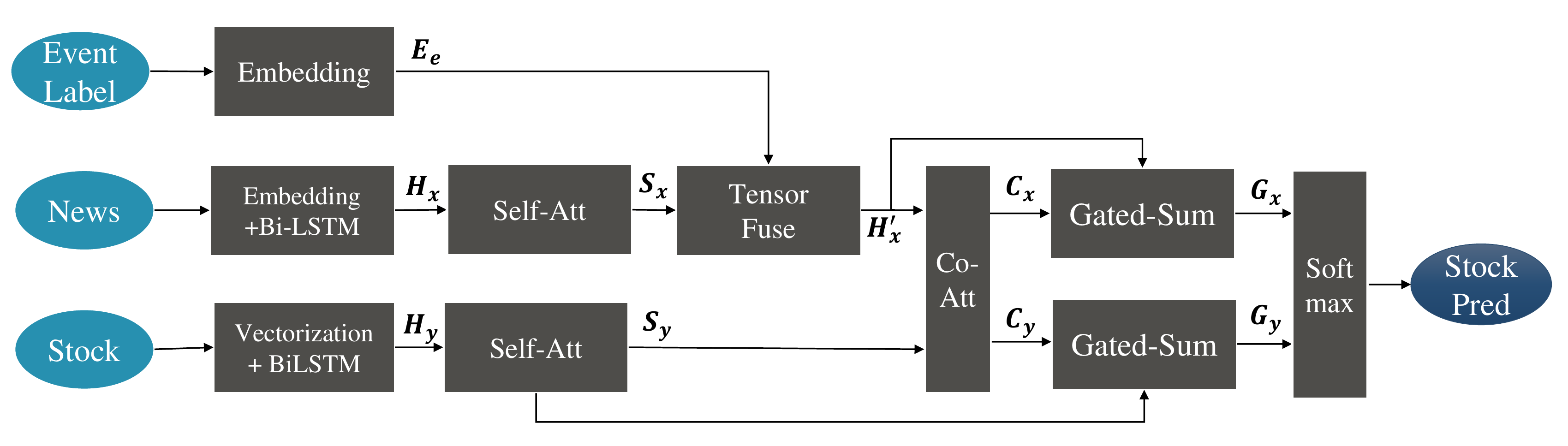}
\caption{The overview of the proposed SSPM model.}
\label{fig3SSPM}
\vspace{-0.1in}
\end{figure*}

\section{Proposed Method}
\subsection{Problem Formulation}
Given $N$ samples in the dataset, and the $i$-th sample $(x^i,y^i,e^i,s^i)$ contains the news text $x^i$, the stock trade data $y^i$ in the day before news happens, the event role label $e^i$ generated in Section~\ref{Section2EventDataGenerate} and stock movement label $s^i$. 
$x^{i}=\left \{ x^i_1,x^i_2,...,x^i_L  \right \}$ is a sequence of words with length of $L$. $e^{i}=\left \{ e^i_1,e^i_2,...,e^i_L  \right \}$ is a sequence of labels indicating the event role of each word in $x^{i}$. $y^{i} = \left \{ y^i_1,y^i_2,...,y^i_M  \right \}$ is a sequence of trade record vectors for each trade minute with length of $M$. $s^i\in \{0,1\}$ is the stock movement label telling whether the stock trade price is up or down at prediction time. 
The stock movement prediction task can be defined as assigning movement label for the news input and trade data input.

\subsection{Trade Data Embedding}
\label{Section1MarketEmbedding}
Different from works of~\citep{Raw4ACL18VAE,Event4MultiModal1IEEE18MultiSource} who use limited daily-level stock trade data (stock close price and daily trade volume, for example), we adopt the minute-level stock data to describe the stock movement in a more detailed way. For each minute when at least one trade happens, we collect the following items: (1) First/last/highest/lowest trade price of the minute; (2) Total trade volume/value of the minute; (3) Volume-weighted average trade price.
The stock trade data is of time series data, so in order to apply the powerful time series neural models, we transfer the raw trade features into trade data embedding $\bm{E_y}$.
The following combination performs best on the develop set:
\begin{itemize}
    \item Raw Number: first/last/highest/lowest trade price, total trade volume and volume-weighted average trade price
    \item Change Rate: change rates of all the raw number items compared to last minute
\end{itemize}
Now we get 12 feature numbers for each trade minute. We set the length of time step to 10 minutes. Then we get the trade data embedding $\bm{E_y} \in \mathbb{R}^{T \times D_s}$. $T = {\rm M/10}$ and ${\rm D_s = 120}$. ${\rm M}$ is the length of the trade minutes.
Finally, we adopt the min-max scale method for each stock's samples and pad the time steps less than 10 minutes with last trade minute's data.

\subsection{Base Model: Structured Stock Prediction Model (SSPM)}
Figure~\ref{fig3SSPM} shows the overview of SSPM. We first transfer various sources of input $(x,y,e)$ into dense vectors. Then we get the representations of text and stock data through bi-directional Long Short Term Memory (BiLSTM) and self-attention. Then we fuse text and event structure to access the structure-aware text representation. Finally, we interact text and stock data by co-attention to predict stock movement.
There are 4 modules in SSPM: input embedding, single-modal information representation, bi-modal information interaction and prediction. Experiment results show that SSPM outperforms all the baselines.

\subsubsection{Input Embedding}
\label{inputEmbedding}
The purpose of this module is to transfer various sources of input ($x,y,e$) into dense vectors. For words in finance news $x$, we use both word-level pretrained embedding Glove~\citep{Glove} and character-level pretrained embedding ELMo~\citep{ELMO} for the purpose of representing words better from different levels. Then we concatenate them to get the final word representation $\bm{E_x} \in \mathbb{R}^{L \times D_w}$.
We use the method proposed in Section ~\ref{Section1MarketEmbedding} to get the stock trade data embedding $\bm{E_y} \in \mathbb{R}^{T \times D_s}$.
Besides, we embed event role labels $e$ into dense vectors $\bm{E_e} \in \mathbb{R}^{L \times D_e}$ using a parameter matrix initialized with random values. $D_w,D_s,D_e$ are the embedding dimensions of word, stock and event role, respectively. $T$ is the length of stock time-steps.

\subsubsection{Single-modal Information Representation}
\label{SingleModalRpresent}
The purpose of this module is to get the representations for both news and stock trade data independently. After accessing the input embedding, we employ BiLSTM to encode the $\bm{E_x}$ and $\bm{E_y}$ :
{\setlength{\abovedisplayskip}{5pt}
\setlength{\belowdisplayskip}{5pt}
\begin{flalign*}
& \bm{H_x} = {\rm BiLSTM_{x}}(\bm{E_x}) \nonumber \\
& \bm{H_y} = {\rm BiLSTM_{y}}(\bm{E_y})  \nonumber 
\end{flalign*}}
Now we access the sentence representation $\bm{H_x}\in \mathbb{R}^{L\times 2h}$ and daily stock trade representation $\bm{H_y}\in \mathbb{R}^{T\times 2h}$. $h$ is the hidden size of BiLSTM.
In order to enhance the learning ability, we use the self-attention to allow the $\bm{H_x}$ and $\bm{H_y}$ to have a look at themselves and make adjustment. We apply the bi-linear attention method which have proven~\citep{Attention1Main,Attention4MMWGT} more powerful in learning ability. Here are the formulas for $H_x$:
{\setlength{\abovedisplayskip}{5pt}
\setlength{\belowdisplayskip}{5pt}
\begin{flalign*}
& W_{SA}^x = {\rm softmax}(\bm{H_x}\cdot W_1 \cdot \bm{H_x^\top})\\
& \bm{S_x} = W_{SA}^x \cdot \bm{H_x} 
\end{flalign*}}
$ W_1$ is a trainable weight matrix and $\bm{S_x} \in \mathbb{R}^{L\times 2h}$. In the same way we get the self-attention result of the stock data: $\bm{S_y} \in \mathbb{R}^{T\times 2h}$. 

In the $<$S,P,O$>$ method, event roles are extracted as separated phrases where some words are ignored and the word order information is missing. Instead, we fuse the text representation $S_x$ with the event role embedding $E_e$ to capture the structure information and remain the word order at the same time. $E_e$ contains both word-level (event role) and sentence-level (BIO label) information, which is similar with $S_x$, so we select to fuse $E_e$ with $S_x$ instead of $E_x$.
Here we adopt the fusion function used in~\citep{Attention1Main,FusionFunction} to fuse the event structure and text effectively:
{\setlength{\abovedisplayskip}{5pt}
\setlength{\belowdisplayskip}{5pt}
\begin{flalign*}
\bm{H'_x}= \sigma(W_f[\bm{S_x};\bm{E_e};\bm{S_x}-\bm{E_e};\bm{S_x}\circ \bm{E_e}]+b_f)
\end{flalign*}}
${\rm ;}$ means tensor connection. We ensure $D_e=2h$ so that $E_e$ has the same dimension with $S_x$. $\circ$ means element-wise multiplication and $\sigma$ is the activation function. $\bm{H'_x} \in \mathbb{R}^{L\times 2h}$ is the structure-aware text representation.
 
\subsubsection{Bi-modal Information Interaction}
In this part we conduct the interaction between the two modal information: finance news of text mode and stock trade data of number mode. These two different modal information are highly relevant: the finance news represents the environment variable and the stock trade data represents history movement. The interaction between them can lead to a better understanding of stock movement.
We use the co-attention to interact the bi-modal information: $\textstyle \bm{H'_x} =  \left \{ h_x^{'1},h_x^{'2},\cdots,h_x^{'L} \right \} $ and $\textstyle \bm{S_y} = \left \{ s_y^{1},s_y^{2},\cdots,s_y^{T} \right \} $. The attention weight is computed by the following function:
{\setlength{\abovedisplayskip}{5pt}
\setlength{\belowdisplayskip}{5pt}
\begin{flalign*}
f_{att}(i,j) = {\rm Relu}(h_x^{'i\top}\, W_2\,  s_y^{j})
\end{flalign*}}
$W_2$ is a trainable weight matrix. We use the softmax function to normalize the attention weight:
{\setlength{\abovedisplayskip}{5pt}
\setlength{\belowdisplayskip}{5pt}
\begin{flalign*}
\alpha_{ij}=\frac{e^{f_{att}(i,j)}}{\sum_{k=1}^{T}e^{f_{att}(i,k)}}; \quad 
\beta_{ij}=\frac{e^{f_{att}(i,j)}}{\sum_{t=1}^{L}e^{f_{att}(t,j)}}
\end{flalign*}}
Finally we get the reconstructed representations:
{\setlength{\abovedisplayskip}{5pt}
\setlength{\belowdisplayskip}{5pt}
\begin{flalign*}
c_x^i = \sum_{j=1}^{T} \alpha_{ij}\,s_y^j ; \qquad
c_y^j = \sum_{i=1}^{L} \beta_{ij}\,h_x^{'i}
\end{flalign*}}
Now we access the reconstructed representations $\bm{C_x} = \left \{ c_x^{1},c_x^{2},\cdots,c_x^{L} \right \} $ and $\bm{C_y} = \left \{ c_y^{1}, c_y^{2},\cdots, c_y^{T} \right \}$ based on the attention to another modal information. 
Then we use the gating mechanism to incorporate the original representation and the corresponding attention result:
{\setlength{\abovedisplayskip}{5pt}
\setlength{\belowdisplayskip}{5pt}
\begin{flalign*}
&  \bm{G_x} = {\rm g}(\bm{H'_x},\bm{C_x)}\cdot \bm{C_x} + ({\rm 1- g}(\bm{H'_x},\bm{C_x}))\cdot \bm{H'_x} \\
&  \bm{G_y} = {\rm g}(\bm{S_y},\bm{C_y)}\cdot \bm{C_y} + ({\rm 1- g}(\bm{S_y},\bm{C_y}))\cdot \bm{S_y} 
\end{flalign*}}
where the $g(,)$ is the gating function and we use the non-linear transformation with $sigmod$ activation function in experiment.
\subsubsection{Prediction}
In this module, we concatenate the $\bm{G_x}$ and $\bm{G_y}$ and predict the stock movement label $\widehat{p}$:
{\setlength{\abovedisplayskip}{5pt}
\setlength{\belowdisplayskip}{5pt}
\begin{flalign*}
& \widehat{p}(s|x,y,e) ={\rm softmax}(W_p[\bm{G_x};\bm{G_y}]+b_p)
\end{flalign*}}

\subsection{Advanced Model: Multi-task Structured Stock Prediction Model (MSSPM)}

\begin{figure*}[t]
\centering
\includegraphics[scale=0.48]{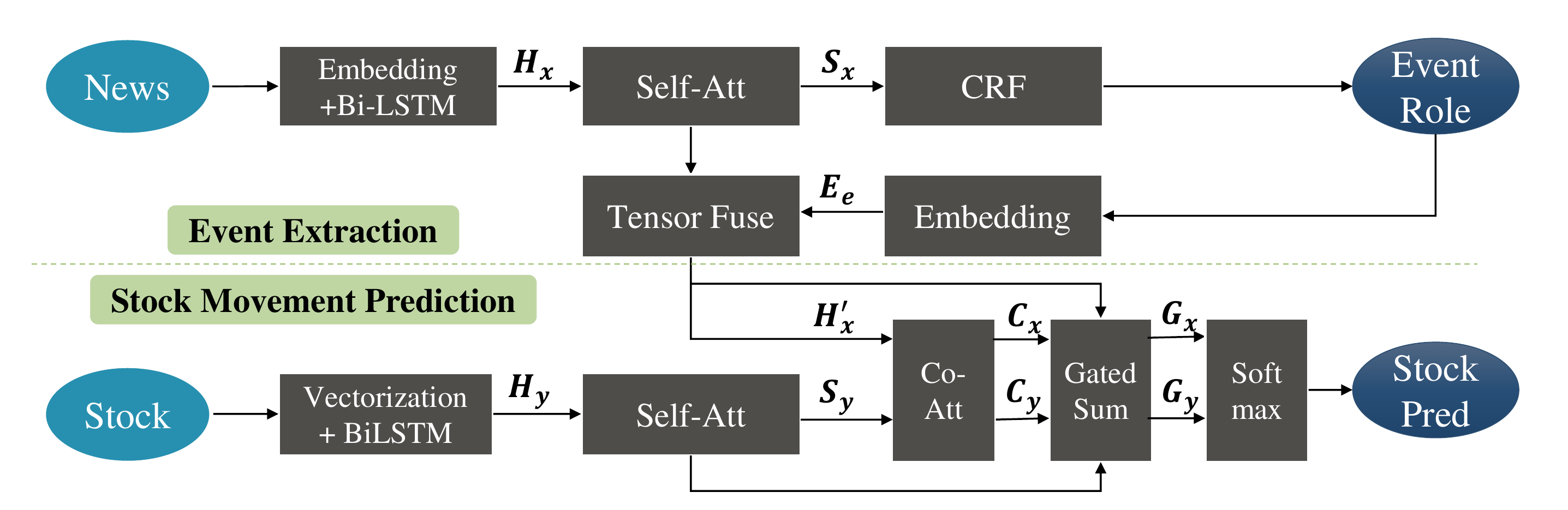}
\caption{The overview of the proposed MSSPM model.}
\label{fig4MSSPM}
\vspace{-0.1in}
\end{figure*}

SSPM can hardly process the uncovered news that can not be recognized by the TFED since the fine-grained event structure information is not provided. MSSPM is designed to handle this issue by employing the generated ${e}$ in Section~\ref{Section2EventDataGenerate} as the distant supervised label to train an event extractor. Furthermore, we design a multi-task framework to jointly learn event extraction and stock prediction because these two tasks are highly related. The quality of event extraction result has a direct influence on the downstream stock prediction task. At the same time, the results of stock prediction can give valuable feedback to event extraction. The multi-task framework can share useful information and make effective interaction between tasks. The overview of MSSPM is shown in Figure~\ref{fig4MSSPM}. The upper half of the dotted line represents the event extraction part. We regard the event extraction task as a sequence labeling task and adopt the self-attended BiLSTM-CRF (conditional random fields) method to make labeling decisions. The lower half stands for the stock movement prediction part which works in a similar way as SSPM.

\subsubsection{Event Extraction}
\label{section_event_extractor}
After accessing the word embedding $\bm{E_x}$, we employ the BiLSTM to get the sentence representation $\bm{H_x}$. Then we employ self-attention to learn a better representation $\bm{S_x}$. Finally we predict the event label and employ CRF to optimize output:
{\setlength{\abovedisplayskip}{5pt}
\setlength{\belowdisplayskip}{5pt}
\begin{flalign*}
& \widehat{e} = {\rm softmax}(W_l\,\bm{S_x} + b_l) \\
& \widehat{e'} = {\rm CRF}(\widehat{e})
\end{flalign*}}
$\widehat{e'}$ is the estimation of the event role. Then we adopt the method introduced in Section~\ref{inputEmbedding} to get the event role embedding $\bm{E_e}$ from $\widehat{e'}$ and adopt the tensor fusion function used in SSPM to get the structure-aware text representation $\bm{H'_x}$.

\subsubsection{Stock Movement Prediction}
The stock movement prediction process is similar with SSPM, and the main difference is the event input $\bm{E_e}$ is predicted from the event extractor in Section~\ref{section_event_extractor}. The stock trade representation $\bm{S_y}$ is accessed in the same way with SSPM. Then we employ the co-attention to interact $\bm{H'_x}$ and $\bm{S_y}$, followed by the gated sum and $softmax$ function to predict the stock movement label $\widehat{s}$.

\subsubsection{Multi-task Learning Object}
The loss function of MSSPM consists of two parts, which are the negative logarithm loss of event extraction and that of stock prediction:
{\setlength{\abovedisplayskip}{5pt}
\setlength{\belowdisplayskip}{5pt}
\begin{flalign*}
& LS_e = -\sum_t e_tlog\,p(e_t|x),\,t=[1,\cdots,L] \\
& LS_s = - s\,{\rm log}\,p(s|x,y)
\end{flalign*}}
We select the weighted sum of these two losses as the final loss of MSSPM:
{\setlength{\abovedisplayskip}{5pt}
\setlength{\belowdisplayskip}{5pt}
\begin{flalign*}
LS = \lambda LS_e/L  + (1-\lambda)LS_s
\end{flalign*}}
The $\lambda$ is a hyper-parameter to balance two losses. $LS_e$ is divided by the number of words to ensure it is comparable with $LS_s$. The experiment result shows that model performs best on develop set when $\lambda=0.43$. 

\section{Experiment}
\label{SectionExperimentData}
The experiment data is from the professional finance news providers Reuters\footnote{Source Reuters News cThomson Reuters cREFINITIV, \url{https://www.thomsonreuters.com/en.html}}.
We collect finance news related to TOPIX top 1000 stocks\footnote{Each news in Reuters has a manual field indicating its related stock(s) and we use it to filter the stock related news.} from 2011 to 2017. The raw data contains both news headline and body, and we use headline only since the headline contains the most valuable information and has less noise than news body. 
We collect stock trade data for news happens in/out of trade time (9:00 AM - 15:00 PM in trade day) differently. For those news happens in trade time, we collect the trade data from 9:00 AM to the last minute before news happens. And for those news happens out of trade time, we collect the trade data of last trade day. We want to ensure no trade data after news happens are included in the input in which situation the market reactions are leaked to the model. We get about 210k data samples finally. 
Following~\citep{DingEvent2,Raw4ACL18VAE}, the stock movement is divided into two categories: stock up/down.
The stock up and down rates are 45$\%$ and 55$\%$ in our dataset, respectively. We adopt TOPIX Sector Index to correct the stock movement in order to eliminate the influence of macro news and the details are introduced in supplementary materials C. In experiment, we reserve 10k samples for developing and 10k samples for testing. 
The samples in train set are previous to samples in valid set and test set to avoid the possible information leakage.
All the rest 190k samples are applied for training SSPM while only the dictionary covered part (about 70$\%$) in the 190k samples are applied for training MSSPM to acquire a high-quality event extractor.  
We tune the hyper-parameters on the development set and test model on the test set. 
The evaluation metrics are accuracy and Matthews Correlation
Coefficient (MCC). MCC is often used in stock forecast~\citep{Raw4ACL18VAE,DingEvent3} because it can overcome the data imbalance issue. More experiment details are listed in the supplementary material D.

\section{Results and Analysis}
We analyze the results of experiments in this section. Firstly, we compare the proposed methods with the baselines in Section~\ref{section_baseline}. Then we analyze the effect of event structure input in SSPM in Section~\ref{Effect of Event Structure} and analyze the MSSPM method in Section~\ref{Analysis of MSSPM}. Lastly, we conduct the ablation study in Section~\ref{Ablation Study}. We also conduct error analysis over different times' news, which is shown in the supplementary material B.
\subsection{Comparison With Baselines}
\label{section_baseline}
The following baselines are used in this work.
\begin{itemize}
    \item \textbf{Bagging Decision Tree:} This method adopt bagging ensemble algorithm to combine 20 Decision Tree classifiers to make the prediction. It outperforms all the other traditional machine learning methods we tried.
    \item \textbf{Sentiment Analysis:} This method~\citep{Raw2Emnlp14sentiment} conducts sentiment analysis on news headlines to predict stock movement.
    \item \textbf{Target Specific Representation:} This method~\citep{Raw1Coling18NewsBody} employs the news headline as the target to summarize the news body in order to utilize the abundant information of news body.
    \item \textbf{Triple Structure:} This method~\citep{DingEvent1} adopts the $<$S,P,O$>$ triple to represent the event structure.
    \item \textbf{Weighted Triple Structure:} This method ~\citep{DingEvent2} adds trainable weight matrices in $<$S,P,O$>$ to enhance fitting ability.
    \item \textbf{Triple Structure with RBM:} This method \citep{Event4MultiModal1IEEE18MultiSource} uses Restricted Boltzmann Machine to handle the $<$S,P,O$>$ and then adopts multi-instance learning to model the latent consistencies of different data sources. Because tweet data are contained in news data\footnote{Some related tweets about stocks are also provided by Reuters mixed with news.} in our dataset, our implementation uses news and stock data instead of news, tweet and stock data.
\end{itemize}

\begin{table}[t]
\centering
\scalebox{0.71}{
    \begin{tabular}{|lccc|}
\hline
\multicolumn{1}{|c}{Method}                                            & \begin{tabular}[c]{@{}c@{}}Event\end{tabular} & Acc(\%) & MCC   \\ \hline
Bagging Decision Tree                                                  & No                                                          & 54.9    & 0.096 \\
\begin{tabular}[c]{@{}l@{}}Sentiment Analysis\\ \citep{Raw2Emnlp14sentiment} \end{tabular}      & No                                                          & 62.8    & 0.253 \\
\begin{tabular}[c]{@{}l@{}}Target Specific Rep.\\ \citep{Raw1Coling18NewsBody} \end{tabular}      & No                                                          & 63.7    & 0.275 \\ \hline
\begin{tabular}[c]{@{}l@{}}Triple Structure\\ \citep{DingEvent1} \end{tabular}          & Coarse-grained                                                         & 63.2    & 0.270 \\
\begin{tabular}[c]{@{}l@{}}Weighted Triple Structure\\ \citep{DingEvent2} \end{tabular} & Coarse-grained                                                       & 63.5    & 0.269 \\
\begin{tabular}[c]{@{}l@{}}Triple Structure with RBM\\ \citep{Event4MultiModal1IEEE18MultiSource} \end{tabular}   & Coarse-grained                                                         & 64.0    & 0.278 \\ \hline
MSSPM(proposed)                                                        & Fine-grained                                                       & 65.7    & 0.315 \\
SSPM(proposed)                                                         & Fine-grained                                                        & 66.4    & 0.330 \\
Ensemble  & Fine-grained & \textbf{67.2} & \textbf{0.348} \\ \hline
\end{tabular}}
\caption{Results on test set compared with baselines; the results in this table and following tables have proven significant with $\rm{p<0.05}$ by student t-test.}
\label{Result1Baseline}
\vspace{-0.05in}
\end{table}
Table~\ref{Result1Baseline} is divided into three parts. 
The three baselines in the top part employ the text directly as model input. These methods totally ignore the structure information of text. The three baselines in the middle part take structure information into account. 
These methods consider $<$S,P,O$>$ event roles in all event types though, they miss some important event roles and describe the event roles in a very rough way. Moreover, the word order information is missing under such settings. 
Both SSPM and MSSPM outperform all the baselines. 
These proposed method incorporate the fine-grained event structure in stock movement prediction. It can extract specific fine-grained event structures for different types of finance events. At the same time, this method remains the original word order through the tensor fusion.
Another advantage of our method is that it applies the stock data embedding method for the minute-level stock trade data and conducts interaction between stock data and news data. 
SSPM performs a little better than MSSPM because SSPM adopts more data for training and the learning of event extraction in MSSPM is not perfect. 
The Ensemble method follows a simple rule to combine the SSPM and MSSPM: The TFED covered news is processed by SSPM and the uncovered news is processed by MSSPM. It achieves the best result among all the baselines and proposed methods.

\begin{table}[t]
\centering
\scalebox{0.85}{
\begin{tabular}{|l|cc|}
\hline
\multicolumn{1}{|c|}{Input Form}   & Acc(\%) & MCC   \\ \hline
No Text         & 58.1             & 0.161          \\ 
No Event (Raw Text)      & 62.2             & 0.246          \\ 
Coarse-grained Event    & 64.6            & 0.291          \\
Fine-grained Event      & \textbf{66.4}    & \textbf{0.330} \\ \hline
\end{tabular}}
\vspace{-0.07in}
\caption{Different Text Input Forms in SSPM.}
\label{table_event}
\vspace{-0.05in}
\end{table}

\subsection{Effect of Event Structure}
\label{Effect of Event Structure}
In this section, we analyze the effect of event structure. Although there have been some comparisons of different event structures in Section~\ref{section_baseline},  the models are different. In this section, we conduct an experiment based on the SSPM model and change different text input forms to check the impact of the event structure.
We design 4 different text input forms for SSPM: (1) No Text method takes no text information as input and relies entirely on trade data to predict stock movement; (2) No Event takes the raw news text as model input and removes the event input from SSPM; (3) Coarse-grained Event employs the coarse-grained event structure $<$S, P, O$>$ as event input of SSPM; (4) Fine-grained Event is the proposed method to utilize the category-specific fine-grained event as model input. The results are shown in Table~\ref{table_event}. 
We can find that all the there methods adding text input outperform the No Text method, which proves the effect of finance news.
Both the Coarse-grained and Fine-grained Event methods bring improvement to the prediction result, which shows that the event structure is very useful. Moreover, the Fine-grained Event method brings larger improvement than the Coarse-grained Event method, which demonstrates that utilizing fine-grained events is more helpful to help model understanding the semantic information of news text.

\subsection{Analysis of MSSPM}
\label{Analysis of MSSPM}
Although SSPM performs well in stock prediction, there are two important issues with it. Firstly, 29$\%$ news in our dataset can not be recognized by TFED, and the Table~\ref{Result2Uncover} shows that the result of uncovered data is obvious lower than the covered data. Secondly, TFED is domain specific, so the generalizability of SSPM may be restricted. MSSPM is designed to handle these two issues.
\begin{table}[t]
\centering
\scalebox{0.95}{
    \begin{tabular}{|c|cc|cc|}
\hline
Data  & \multicolumn{2}{c|}{Covered}   & \multicolumn{2}{c|}{Uncovered}       \\ \hline
Metric & Acc($\%$)  & MCC  & Acc($\%$)  & MCC              \\ \hline
SSPM  & 67.6 & 0.351 & 63.4          & 0.267          \\ 
MSSPM & 65.9          & 0.318          & \textbf{65.2} & \textbf{0.305}     \\ \hline
\end{tabular}}
\vspace{-0.05in}
\caption{Result on different sets of test data. The covered set means samples recognized by the TFED and the uncovered set means the samples out of the dictionary. These two sets account for around 30$\%$ and 70$\%$ of the test set, respectively.}
\label{Result2Uncover}
\end{table}

\begin{table}[t]
\centering
\scalebox{0.88}{
    \begin{tabular}{|c|c|cc|}
\hline
Task       & Event Extraction & \multicolumn{2}{c|}{Stock Prediction} \\ \hline
Metric     & Micro-F1(\%)     & Acc(\%)           & MCC               \\ \hline
Pipeline   & 79.2          & 64.8              & 0.297             \\ 
Multi-Task & \textbf{84.3}  & \textbf{65.7}     & \textbf{0.315}    \\ \hline
\end{tabular}}
\caption{Comparison of pipeline method and multi-task learning method (MSSPM). The pipeline method trains the event extractor first and then predicts the stock. We report the micro-F1 score for the event extraction task.}
\label{Result3Pipeline}
\vspace{-0.2in}
\end{table}
As shown in Table~\ref{Result2Uncover}, although the performance of MSSPM is lower than SSPM on the covered test set, its performance is higher than SSPM on the uncovered test set.
The performance decrease of MSSPM after transferring from covered set to uncovered set is much smaller than SSPM's, which proves MSSPM has higher transferability. The uncovered news can be regarded as events of new types, and MSSPM performs better on it by learning event extraction, which improves the generalizability of the structured stock prediction method. 
As shown in Table~\ref{Result3Pipeline}, the performance of multi-task learning is clearly better than the pipeline method, which confirms our assumption that these two tasks are highly related and the joint learning improves both of their results.

\subsection{Ablation Study}
\label{Ablation Study}
In this section, we report and analyze the results of ablation study. We remove different components of both SSPM and MSSPM to check their effect.
As shown in Table~\ref{Result4Ablation1} and Table~\ref{Result5Ablation2}, we found that the model performance drops in all the ablation experiments as expected. The fusion function, attention mechanism (both self-attention and co-attention) and the gating mechanism are all helpful for both SSPM and MSSPM.  We can observe an obvious decrease after removing fusion function (adopt adding method instead) both in SSPM ($\downarrow 1.5$ of Acc) and MSSPM ($\downarrow 1.1$ of Acc), which demonstrates that the fusion function combines the event structure and the news text effectively. Besides, the co-attention between news and stock trade data also plays an important role in both models.

\begin{table}[t]
\centering
\scalebox{0.85}{
    \begin{tabular}{|lcc|}
\hline
\multicolumn{1}{|c}{Metric}            & Acc(\%)       & MCC            \\ \hline
\multicolumn{1}{|c}{SSPM}         & \textbf{66.4} & \textbf{0.330} \\ \hline
\emph{\rm w/o Fusion Function}   & 64.9($\downarrow 1.5$)     & 0.298($\downarrow 0.032$)   \\
\emph{\rm w/o Self-Attention}  & 66.0($\downarrow 0.4$)     & 0.319($\downarrow 0.011$)   \\
\emph{\rm w/o Co-Attention} & 65.1($\downarrow 1.3$)     & 0.303($\downarrow 0.027$)   \\
\emph{\rm w/o Gated Sum} & 65.6($\downarrow 0.8$)     & 0.314($\downarrow 0.016$)   \\ \hline
\end{tabular}}
\vspace{-0.07in}
\caption{Ablation Study of SSPM. }
\label{Result4Ablation1}
\vspace{-0.05in}
\end{table}

\begin{table}[t]
\centering
\scalebox{0.85}{
    \begin{tabular}{|lcc|}
\hline
\multicolumn{1}{|c}{Metric}               & Acc(\%)       & MCC            \\ \hline
\multicolumn{1}{|c}{MSSPM}         & \textbf{65.7} & \textbf{0.315} \\ \hline
\emph{\rm w/o Fusion Function}   & 64.6($\downarrow 1.1$)     & 0.292($\downarrow 0.023$)   \\
\emph{\rm w/o Self-Attention}   & 65.5($\downarrow 0.2$)     & 0.310($\downarrow 0.005$)   \\
\emph{\rm w/o Co-Attention}  & 64.9($\downarrow 0.8$)     & 0.298($\downarrow 0.017$)   \\
\emph{\rm w/o Gated Sum}  & 65.2($\downarrow 0.5$)     & 0.304($\downarrow 0.011$)   \\
\emph{\rm w/o CRF}              & 64.2($\downarrow 1.5$)     & 0.285($\downarrow 0.030$)   \\ \hline
\end{tabular}}
\vspace{-0.07in}
\caption{Ablation Study of MSSPM.}
\label{Result5Ablation2}
\vspace{-0.2in}
\end{table}

\section{Conclusion}
In this work, we propose to incorporate the fine-grained events in stock movement prediction task. We propose the TOPIX Finance Event Dictionary with domain experts' knowledge and extract fine-grained events automatically. We propose SSPM to incorporate fine-grained events in stock movement prediction which outperforms all the baselines. Besides, to handle the uncovered news, we use the event data as the distant supervised label to train a multi-task framework MSSPM. The results show that MSSPM performs better on uncovered news and improves the generalizability of the structured stock prediction method.
\section{Acknowledgement}
This work is supported by a Research Grant from Mizuho Securities Co., Ltd. Three experienced stock traders from Mizuho Securities provide the professional support for the TFED dictionary. Mizuho Securities also provide experiment data.

\bibliography{emnlp-ijcnlp-2019}
\bibliographystyle{acl_natbib}

\appendix

\section{TOPIX Finance Event Dictionary}
\label{supplementary materials A}

Table 7 shows the TOPIX Finance Event Dictionary we used in this paper for generating fine-grained events automatically. Table 1 displays all the 32 event types in 8 categories as well as their trigger words and event roles.
\label{sec:supplemental}

\begin{table*}[p]
\centering
\scalebox{0.85}{
    \begin{tabular}{|l|l|l|l|}
\hline
\textbf{Category}                                              & \textbf{Event}     & \textbf{Trigger Word(s)}                                                                                                   & \textbf{Event Role}                                                                                                    \\ \hline
Affairs                                                     & Announcement          & announcement,announce                                                                                                  & Who Content Time                                                                                                             \\ \cline{2-4} 
                                                            & Legal Issues          & court,litigation,lawsuit                                                                                                      & \begin{tabular}[c]{@{}l@{}}Who  Target Reason \\ Location Requirement\end{tabular}                                        \\ \cline{2-4} 
                                                            & Personnel Affairs     & appint,name...as                                                                                                       & Who Title Person                                                                                                         \\ \cline{2-4} 
                                                            & Recall                & recall                                                                                                                 & Firm Amount Target Time	Location                                                                                         \\ \hline
Business                                                    & Buy                   & buy                                                                                                                    & Firm Target Price Time                                                                                                   \\ \cline{2-4} 
                                                            & Cooperation           & alliance,partnership                                                                                                   & Firm With-Firm Filed                                                                                                     \\ \cline{2-4} 
                                                            & Deals                 & order,deal,trade                                                                                                       & Firm Provider Product Value Number                                                                                                      \\ \cline{2-4} 
                                                            & Demand/Supply         & demand,supply                                                                                                          & Who Type Content Time                                                                                                    \\ \cline{2-4} 
                                                            & Investment            & \begin{tabular}[c]{@{}l@{}}take...charge,invest,\\ investment\end{tabular}                                             & Firm Target Money                                                                                                        \\ \cline{2-4} 
                                                            & Sales                 & sale                                                                                                                   & \begin{tabular}[c]{@{}l@{}}Firm Result-number Result-rate \\ Comparason Direction\\ Location Time\end{tabular}           \\ \hline
\begin{tabular}[c]{@{}l@{}}Corporate \\ Action\end{tabular} & Bond Issurance        & share issuance                                                                                                         & Firm Time Type How                                                                                                       \\ \cline{2-4} 
                                                            & Dividend              & div,dividend                                                                                                           & Firm Type Time                                                                                                           \\ \cline{2-4} 
                                                            & Fundings              & fund,funding                                                                                                           & Who Action Target Time Location                                                                                          \\ \cline{2-4} 
                                                            & IPO                   & ipo                                                                                                                    & Firm Market IBD Value Time                                                                                               \\ \cline{2-4} 
                                                            & Joint venture         & jv,joint venture                                                                                                       & \begin{tabular}[c]{@{}l@{}}Firm Target-Firm \\ Target-Rate Time\end{tabular}                                             \\ \cline{2-4} 
                                                            & M\&A                  & acquisition,merge,acquire                                                                                              & Firm Time Method                                                                                                         \\ \cline{2-4} 
                                                            & Share Buyback         & buy...back,share repurchase                                                                                            & Who Share-number Money                                                                                                   \\ \cline{2-4} 
                                                            & Share Issurance       & share issuance                                                                                                         & Who Money Purpose                                                                                                        \\ \cline{2-4} 
                                                            & Stock Split           & stock split                                                                                                            & \begin{tabular}[c]{@{}l@{}}Firm Before-Price After-Price \\ Change-Rate Time\end{tabular}                                \\ \cline{2-4} 
                                                            & Projects\&Productions & product,production,project                                                                                             & Firm Projects/Products Time                                                                                              \\ \hline
Eearnings                                                   & Eearnings Profit      & \begin{tabular}[c]{@{}l@{}}profit,group result,earnings\\ parent result,financial result,\end{tabular}                 & \begin{tabular}[c]{@{}l@{}}Firm Type Value Change-Rate \\ Time \end{tabular}                                         \\ \cline{2-4} 
                                                            & Eearnings Adjustment  & earnings adjust,profit adjust                                                                                          & \begin{tabular}[c]{@{}l@{}}Firm Before-Value After-Value\\  Reason\end{tabular}                                \\ \cline{2-4} 
                                                            & Eearnings Forecast    & group forecast,parent forecast                                                                                         & \begin{tabular}[c]{@{}l@{}}Firm Value Change-Rate \\  Reason Time\end{tabular}                                \\ \hline
ESG                                                         & Energy                & solar power,plant                                                                                                      & Who Type Movement Location                                                                                               \\ \cline{2-4} 
                                                            & Social                & social,socirty                                                                                                         & Who Action Purpose Time                                                                                                  \\ \cline{2-4} 
                                                            & Government            & gove,government                                                                                                        & Country Action Target Time                                                                                               \\ \hline
Market                                                      & Market Movement       & \begin{tabular}[c]{@{}l@{}}shares down ,shares up,\\ trade hult\end{tabular}                                         & Firm Direction Movement                                                                                                  \\ \cline{2-4} 
                                                            & Share Holders Acion   & holder,investor                                                                                                        & Firm Action Reason Time                                                                                                  \\ \hline
Other                                                       & Comodities            & \begin{tabular}[c]{@{}l@{}}oil ,coal,gassteel ,fuel\\ crude, copper\end{tabular}                                       & \begin{tabular}[c]{@{}l@{}}Comodity Markey Before-Price \\ After-Price Change-Direction \\ Change-Rate Time\end{tabular} \\ \cline{2-4} 
                                                            & Oversea               & \begin{tabular}[c]{@{}l@{}}u.s.,us,american,china,uk,eu,\\ europe,aisa,asian,indonesia,\\ india,australia\end{tabular} & Region Action Time                                                                                                       \\ \hline
Ratings                                                     & Broker Ratings        & \begin{tabular}[c]{@{}l@{}}target price,rating \\ raise...price\\ swhitch...to,scah...to\end{tabular}                  & \begin{tabular}[c]{@{}l@{}}Reviewer Target-Price Directtion\\ FirmNew Old-Price Change-Rate \end{tabular}               \\ \cline{2-4} 
                                                            & Credit Ratings        & moody                                                                                                                  & \begin{tabular}[c]{@{}l@{}}Reviewer New-Rating Before-Rating\\ Target Direction\end{tabular}                            \\ \hline
\end{tabular}}
\label{TOPIX Finance Event Dictionary}
\caption{TOPIX Finance Event Dictionary}
\end{table*}

\section{Error Analysis}

\begin{figure*}[t]
\centering
\includegraphics[scale=0.25]{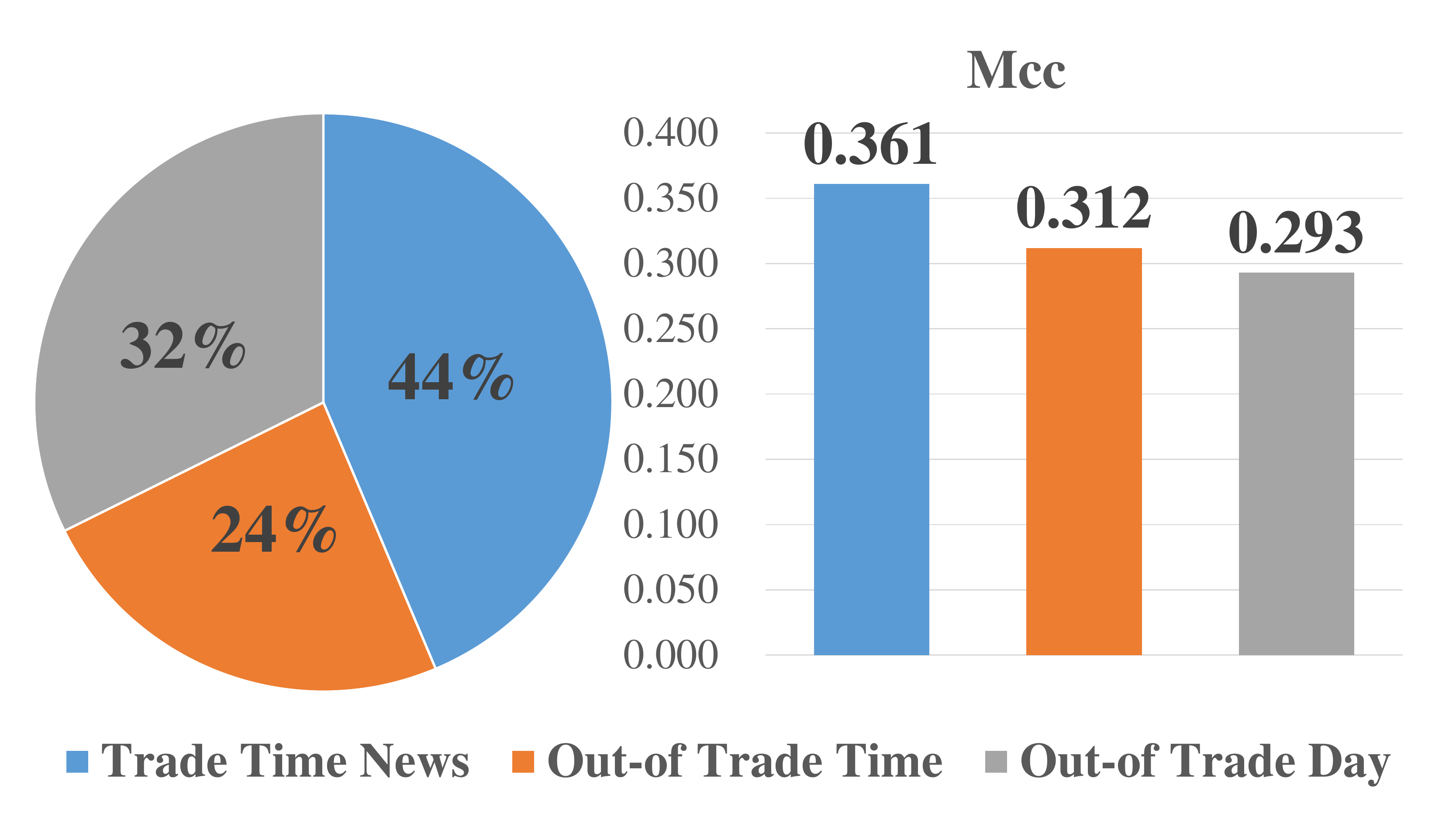}
\caption{The percentage (all dataset) and performance (SSPM on test set) of different times' news.}
\label{fig1ErrorPart}
\end{figure*}

We observe that there are significant gaps in the model performance on different times' news. The Figure~\ref{fig1ErrorPart} shows the percentage of the three different times' news in dataset. The trade time news (44$\%$) means news happens in trade time (9:00 AM - 15:00 PM in trade day); the out of trade time news (24$\%$) represents the news that takes place in trade day, but not in trade time; the out of trade day news (32$\%$) means the news happens in weekend or holiday. 
The experiment result shows that SSPM performs best on trade time news (Acc:68.0$\%$, Mcc:0.361) followed by out of trade time news (Acc:65.8$\%$, Mcc:0.312). SSPM performs worst on out of trade day news (Acc:64.6$\%$, Mcc: 0.293). We summarize this result as the closer the news to trade time, the better our method works. It can be explained that the trade time news have the most direct influence on the stock trading because traders can make immediate reactions to the released news. As for the out of trade time news and the out of trade day news, stock traders can not make immediate reactions and their attitudes may be influenced by other factors, such as the trend of overseas exchanges and the movement of the commodities. The out of trade day news is further away from trade time compared to out of trade time news, so there are more uncertain factors which result in the decrease of model performance.

\section{Stock Movement Label Set}
For those samples in trade time, we compare the stock close price with the price of the minute news happens; for those samples out of trade time, we compare the next trade day's open price with the last trade day's close price. We only use the micro news (specific stock related news) and ignore the macro news, so we use the stock related TOPIX sector index to correct the stock change rate:
{\setlength{\abovedisplayskip}{5pt}
\setlength{\belowdisplayskip}{5pt}
\begin{flalign}
& R_f = R_{cmp} - R_{sec} \nonumber
\end{flalign}}
$R_f$ means the final change rate; $R_{cmp}$ means the raw company stock change rate and $R_{sec}$ means the stock related sector index change rate. For example, for Toyota's stock movement, we use the TPTRAN Index\footnote{TOPIX Transport Index, which indicate the overall trend of the transport market.} which Toyota belongs to for correction. Then the sample is given an up/down label depends on the $R_f$ is positive/negative. 

All the sector indexs used in TOPIX are listed here: TPTRAN Index, TPNBNK Index, TPELMH Index, TPCOMM Index, TPWSAL Index, TPLAND Index, TPSERV Index, TPNCHM Index, TPRETL Index, TPPROD Index, TPINSU Index, TPMACH Index, TPFOOD Index, TPPHRM Index, TPREAL Index, TPRUBB Index, TPPREC Index, TPFINC Index, TPOIL Index, TPCONT Index, TPIRON Index, TPSECR Index, TPELEC Index, TPAIR Index, TPMINN Index, TPTEXT Index, TPNMET Index, TPGLAS Index, TPPAPR Index, TPMETL Index, TPMART Index, TPWARE Index, TPFISH Index.

\section{Experiment Details}
The word vocabulary is extracted from the training set and we keep the top 50000 most frequent words. We use the pretrained ELMo\footnote{\url{https://github.com/allenai/allennlp}} embedding with dimension of 512 and Glove pretrained embedding\footnote{\url{https://nlp.stanford.edu/projects/glove/}} with dimension of 300. We fine-tune the embedding weight in the training process. All the LSTMs~\citep{LSTM} use 3 bi-directional layers and the hidden size is 256. We use the dropout regularization with the dropout probability of $0.2$ to reduce overfitting. We use the Adam~\citep{Adam} optimizer with the initial learning rate of 0.001 and the learning decay rate of 0.0005. The batch size is 128. The training epoch is 60 and we use early stop. We conduct all the experiments on 4 Nvidia Titan P100 GPUs.

We regard news before 9:10 AM as out of trade news to avoid that there is no input for trade data
Besides, we regrade news after 14:50 PM as out of trade news to avoid that the input trade data is too close to the compared price. As for the minute that no trade happens, we pad it with last trade minute's value.

We do not study the multi-news for days on end, which are studied in~\citep{Raw4ACL18VAE,HANN} because we find the situation that news about same stock happens in several continuous days is very sparse in real data.

We make daily prediction in this work because we find stock movement is most sensitive to daily stock trade data. Besides, daily prediction is most valuable for stock investment because clients usually use daily stock data to evaluate the performance of stocks. 

We observe that one piece of news may be related to more than one stocks, so we match this kind of news with each related stock and get multi-samples. For the circumstance that there are more than one news occuring in one trade day about the same stock, which occupies a very small proportion (1.4$\%$) in all the data samples, we connect all the news together as input.

\end{document}